\documentclass[aps,prb,twocolumn,groupedaddress,amssymb,amsmath,showpacs]{revtex4}
\usepackage{dcolumn}
\usepackage{graphicx}

\begin{document}


\title{Magnetism in graphene induced by single-atom defects}

\author{Oleg V. Yazyev}
\email[Electronic address: ]{oleg.yazyev@epfl.ch}
\author{Lothar Helm}
\affiliation{Ecole Polytechnique F\'ed\'erale de Lausanne (EPFL), \\
            Institute of Chemical Sciences and Engineering, \\
            CH-1015 Lausanne, Switzerland} 
	      
\date{\today}

\begin{abstract}

We study from first principles the magnetism in graphene induced by 
single carbon atom defects. 
For two types of defects considered in our study, 
the hydrogen chemisorption defect and the vacancy defect, 
the itinerant magnetism due to the defect-induced extended 
states has been observed.
Calculated magnetic moments are equal to 1~$\mu_B$ per hydrogen chemisorption defect
and 1.12$-$1.53~$\mu_B$ per vacancy defect depending on the defect concentration.
The coupling between the magnetic moments is either ferromagnetic or 
antiferromagnetic, depending on whether the defects correspond to the same or to 
different hexagonal sublattices of the graphene lattice, respectively. 
The relevance of itinerant magnetism in graphene to the high-$T_C$ 
magnetic ordering is discussed.

\end{abstract}


\maketitle

\section{INTRODUCTION}

The last two decades were marked with the discoveries of 
new allotropic modifications of carbon and related nanostructures. 
Graphene, the single two-dimensional sheet of 
graphite, is the starting point for many carbon 
nanomaterials, which are commonly called nanographites. 
These materials, being diverse in atomic structure, 
display 
a wide range of electronic properties. 
Magnetism of carbon materials \cite{Makarova05} is of particular 
interest since in current technological applications 
magnetic materials are based on \textit{d} and \textit{f} 
elements. New carbon-based magnetic materials would greatly extend
the limits of technologies relying on magnetism. 
Even more promising is the application of such materials 
in the design of nanoscale magnetic and spin electronics devices.  

While ideal graphite and carbon nanotubes are in itself nonmagnetic, 
experimental observations of magnetic ordering are often explained by the
presence of impurities \cite{Coey02}, boundaries \cite{Okada01,Lee05} or defects \cite{Kim03,Lehtinen04,Duplock04,Carlsson06}. 
Defects in nanographites \cite{Charlier02} 
can be created intentionally by irradiating material 
with electrons or ions \cite{Ruffieux00,Esquinazi03,Hashimoto04,Urita05}. 
By manipulating the conditions of irradiation it is possible to tune, in a flexible way, the properties of the carbon-based materials 
\cite{Banhart99, Miko03,Han03,Kis04}. 
Examples of simple defects in nanographites are single atom 
vacancies and hydrogen chemisorption defects. 
The former defect type is produced upon the irradiation with high energy
particles \cite{Lehtinen04} while the latter is 
the major outcome of the hydrogen plasma treatment \cite{Ruffieux00}. 
The common feature of both types of defects is that only one carbon
atom is removed from the $\pi$ conjugation network of the graphene sheet.
The single-atom defects on the graphene lattice give rise to quasilocalized 
states at the Fermi level \cite{Mizes89,Ruffieux00,Ruffieux05,Pereira06}. 
The graphene lattice is the bipartite lattice. It
can be viewed as two interpenetrating 
hexagonal sublattices of carbon atoms (labeled $\alpha$ and $\beta$). 
When a defect is created in the $\alpha$ lattice, only the $p_z$ orbitals of 
carbon atoms in the $\beta$ sublattice contribute to the quasilocalized 
state, and vice versa. 
These states extend over several nanometers around the defects 
forming characteristic $(\sqrt{3}$$\times$$\sqrt{3})R30^\circ$ superstructures 
recognized in STM images. 
Analyzing the position and the orientation of the  
superstructures one can precisely locate the defect 
and determine the sublattice to which it belongs \cite{Mizes89,Kelly98}.
The fact that quasilocalized states lie at the Fermi level suggests that 
itinerant (Stoner) magnetism can be induced by the electron exchange instability. 
It has been argued recently, that the Stoner ferromagnetism with high Curie
temperatures, $T_C$, can be expected for $sp$ electron systems.\cite{Edwards06} On the other hand,
the Ruderman-Kittel-Kasuya-Yoshida (RKKY) coupling\cite{RKKY} 
of localized magnetic moments in graphene is too weak 
to result in high $T_C$.\cite{Vozmediano05}
Another recent work has shown that the RKKY-type interactions in graphene are suppressed.\cite{Dugaev06}

In this work we investigate using first principles approaches 
the itinerant magnetism originating from the 
quasilocalized states induced by 
single-atom point defects in graphene. 
The results obtained can eventually be extended,
with some precautions, to defects in other nanographites.

\section{COMPUTATIONAL METHODS AND MODELS}

\begin{figure}
\includegraphics[width=8.5cm]{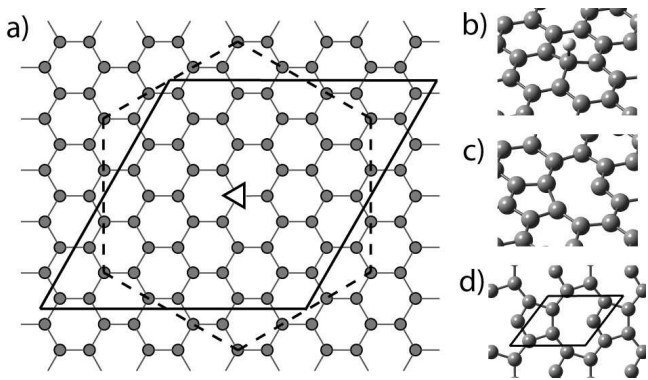}
\caption{\label{fig:eps1}  
(a) Definition of the extendable two-dimensional hexagonal 
lattice of defects in the graphene sheet.  The unit cell and the 
Voronoi cell are shown as full and dashed lines, respectively.
The defective atom is labeled by the triangle according to the 
orientation of the defect state $\sqrt{3}$$\times$$\sqrt{3}$ superstructure.
The size of the supercell shown here corresponds to $9a_{cc}$ 
separation between neighbor defects ($n$=3). 
(b) Structure of the hydrogen chemisorption defect. 
(c) Structure of the vacancy defect. 
(d) Hexagonal closest packing ($n$=1) of vacancy defects with 
the corresponding unit cell.}
\end{figure}

The model system consists of a periodic two-dimensional 
superlattice of defects in graphene (Fig.~\ref{fig:eps1}a). 
The supercell size can be varied resulting in different distances 
$d$ between the neighbor defects on the superlattice 
and, thus, in different defect concentrations. 
The results can be further extrapolated to the cases of low 
defect concentrations. For the chosen supercell, the resulting 
distance between neighbor 
defects is about $3na_{cc}$ where $a_{cc}$=1.42~\AA\ is the 
carbon-carbon distance in graphene. The corresponding number 
of carbon atoms per unit cell is 
$6n^2$. Our investigation is restricted to the cases with $n$=2$-$6. 
The largest system considered ($n$=6)
is characterized by about 25~\AA\ separation between neighbor 
defects, which corresponds to a defect concentration of 0.5\%.

Density functional theory calculations were 
performed using the \texttt{SIESTA} code.\cite{SIESTA,Soler02}
The generalized gradient approximation exchange-correlation density 
functional of Perdew, Burke and Ernzerhof (PBE) \cite{Perdew96} 
was employed. All calculations were performed
in the spin-unrestricted manner using the diagonalization-based 
method for solving Kohn-Sham
equations. The shifted Monkhorst-Pack grids \cite{Monkhorst76} 
corresponding to a cutoff of 100 Bohr were used to sample the Brillouin 
zone in two dimensions. Atomic positions and cell dimensions were relaxed. 
The numerical atomic orbital basis set of single-$\zeta$ plus one 
polarization function (SZP) quality was used for the 
whole range of models studied. All calculations for the models 
with $n$=2$-$4 were reproduced using the basis set of 
double-$\zeta$ plus one polarization function (DZP) quality. 
For all electronic structure quantities discussed in this study
(magnetic moment, Fermi levels and band maxima), there 
is a good agreement between the results of the two basis 
sets, despites the slight overestimation of the C$-$C bond 
length found in the SZP calculations.

\section{DISCUSSION OF RESULTS}

In the following we present our results for the two types 
of defects mentioned above. The structure of the hydrogen
chemisorption defect is shown in Fig.~\ref{fig:eps1}b. This defect 
is characterized by the slight protrusion of the hydrogenated carbon 
atom and the very small displacement of all other neighbor 
carbon atoms.\cite{Ruffieux02,Duplock04}
The single atom vacancy defect in graphene
is nearly planar (Fig.~\ref{fig:eps1}c). The local three-fold symmetry 
breaks down due to the Jahn-Teller distortion induced by the
reconstruction of two dangling bonds left after removing the carbon atom. 
This gives rise to the in-plane displacement of other carbon 
atoms in the graphene lattice.\cite{Lu04,Lehtinen04}
The third dangling bond is left unsaturated providing 
a contribution of magnitude 1~$\mu_B$ to the intrinsic 
magnetic moment of the defect. 
For the case of the vacancy type defect (Fig.~\ref{fig:eps1}d) 
in the closest packing geometry ($n$=1) no single 
six-membered ring remains. This interesting structure can be considered 
as yet another hypothetical allotropic 
modification of carbon for which one may expect a high specific magnetic moment.   

\begin{figure}
\includegraphics[width=7.5cm]{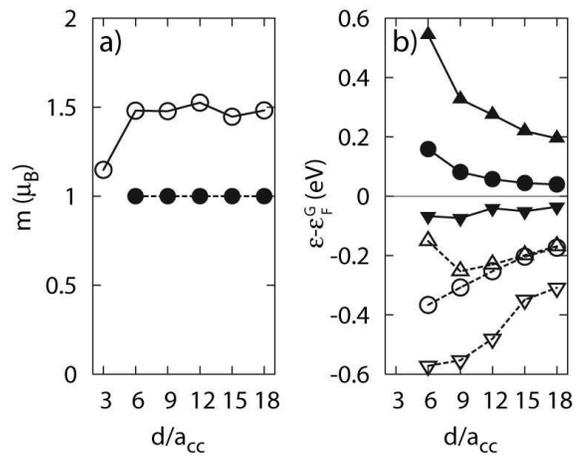}
\caption{\label{fig:eps2}  (a) Calculated 
magnetic moment per defect vs. the separation $d$ between the 
neighbor hydrogen chemisorption defects ($\bullet$) 
and the neighbor vacancy defects ($\circ$). 
(b) Fermi energies ($\bullet$,$\circ$), majority
spin ($\blacktriangle$,$\triangle$) and minority spin ($\blacktriangledown$,$\triangledown$) bands maxima versus 
the defect separation $d$ for the hydrogen chemisorption 
defects (filled symbols) and the vacancy defects (open symbols).
The Fermi level of ideal graphene is taken as zero.}
\end{figure}

\begin{figure}
\includegraphics[width=8.5cm]{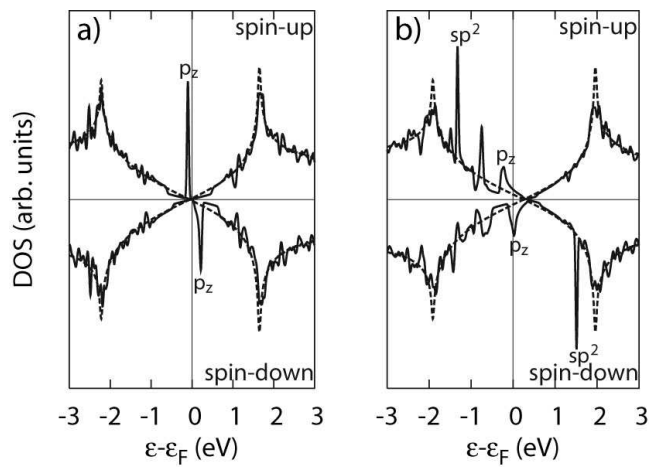}
\caption{\label{fig:eps3} Density of states plots for the systems 
with the hydrogen chemisorption defects (a) and 
with the vacancy defects (b) ($n=4$). The dashed line shows the 
density of states of the ideal graphene. Labels indicate the 
character of the defect states.}
\end{figure}

\begin{figure}
\includegraphics[width=8.5cm]{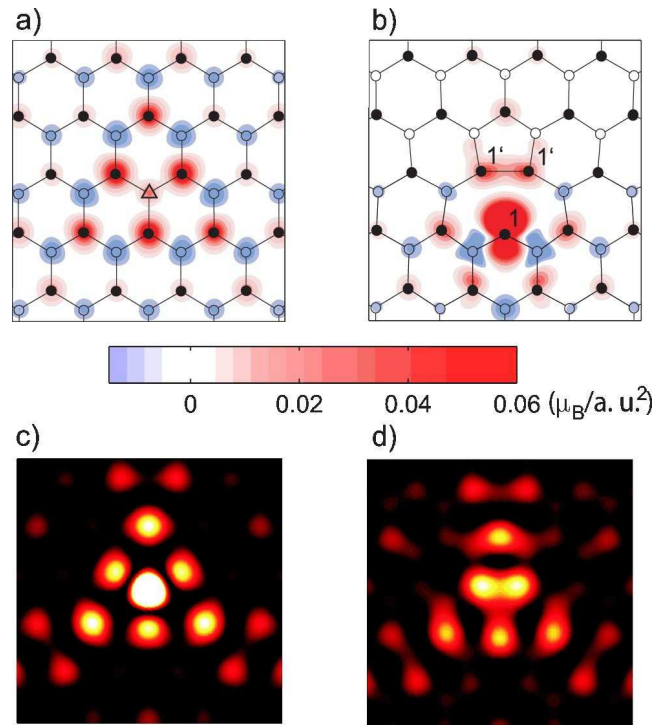}
\caption{\label{fig:eps4} (Color online) Spin density projection (in $\mu_B$/a.u.$^{2}$) 
on the graphene plane around (a) the hydrogen chemisorption defect 
($\triangle$) and (b) the vacancy defect in the $\alpha$ sublattice. 
Carbon atoms corresponding to the $\alpha$ sublattice ($\circ$) and 
to the $\beta$ sublattice ($\bullet$) are distinguished. Simulated 
STM images of the defects are shown in (c) and (d), respectively. }
\end{figure}

Magnetism induced by the presence of quasilocalized 
defect states $\psi_d(\vec{r})$  has been observed in the case 
of both defect types.
The hydrogen chemisorption defect gives rise to the strong 
Stoner ferromagnetism \cite{Mohn03} with a magnetic moment of 
1~$\mu_B$ per defect at 
all studied concentrations (Fig.~\ref{fig:eps2}a). 
The flat defect bands give rise to the very narrow peaks ($W$$<$0.2~eV)
which are necessary for the stability of magnetic ordering at high
temperatures.\cite{Edwards06}
The defect band maxima for the majority spin and the minority 
spin components lie, respectively, lower and higher than the 
the Fermi levels for both defective and ideal graphene (Fig.~\ref{fig:eps2}b).
The hydrogen chemisorption
motif is charge neutral and spin-polarized in 
the wide range of defect concentrations.
On the contrary, fractional magnetic moments and weak Stoner
ferromagnetism \cite{Mohn03} have been observed for the vacancy-type 
defect models. 
A magnetic moment
of 1.15~$\mu_B$ has been predicted for the closest packing of 
vacancy type defects ($n$=1) (Fig.~\ref{fig:eps1}d), while for 
smaller defect concentrations the magnetic moment 
was found to vary in the range of 1.45$-$1.53~$\mu_B$ per defect (Fig.~\ref{fig:eps2}a). 
For the vacancy-type defect, the total magnetic moment 
is determined by the contribution (1~$\mu_B$) of the localized 
$sp^2$ dangling bond state (atom 1 in Fig.~\ref{fig:eps4})
and the contribution ($<$1~$\mu_B$) of the extended defect 
state, $\psi_d(\vec{r})$ (labeled $p_z$ in Fig.~\ref{fig:eps3}). 

The width of the defect state bands 
and the overall modification of the band structure are larger 
in the case of the vacancy type defects (Fig.~\ref{fig:eps3}b).
The partial spin polarization of $\psi_d(\vec{r})$ (filled majority spin band
and half-filled minority spin band) is explained 
by the self-doping (charge transfer from the bulk to $\psi_d(\vec{r})$), 
which arise from the stabilization of the defect state. 
The stabilization of vacancy defect extended states is possible in 
the case of a significant coupling between the 
second nearest neighbor atoms belonging to the same 
sublattice.\cite{Pereira06}
In the case of the vacancy defect, the indirect coupling 
is justified by the formation of the covalent bond between 
the two carbon atoms 1' (Fig.~\ref{fig:eps4}b) that follows the 
defect reconstruction. 
No such bond is possible in the case of hydrogen chemisorption. 
Thus, the character of the defect-induced magnetism 
depends on the possibility of covalent bonding between the second nearest neighbor atoms
due to the reconstruction. This provides an interesting opportunity
for tailoring magnetic properties of materials. 

The defect state exchange splitting, defined as the difference 
between the corresponding majority spin and minority spin band maxima, 
decreases as the defect concentration decreases. This is not 
surprising since the degree of the localization of the defect 
states depends on the defect concentration.\cite{Pereira06}  
At the lowest studied defect concentration of 0.5\%, the exchange 
splitting $d\epsilon_x$ were found to be 0.23~eV and 0.14~eV for the hydrogen 
chemisorption and vacancy defects, respectively.
In the latter case, the splitting is smaller due to the partial 
spin polarization of the defect band. 
Since in both cases $d\epsilon_x$$>$$k_B T$ for $T\approx$300~K, 
the Stoner theory predicts $T_C$ above  room temperature for defect 
concentrations of the order of 1\%.
The decrease of the Stoner theory $T_C$ due to spin wave excitations
is expected to be ineffective for the case of carbon based materials.\cite{Edwards06} 
At low concentrations the magnetism in defective nanographites 
is expected to be sensitive to the variations of the Fermi energy 
resulting from self-doping, presence of other defects or applied 
bias, and to the disorder-induced broadening.

\begin{figure}
\includegraphics[width=7.5cm]{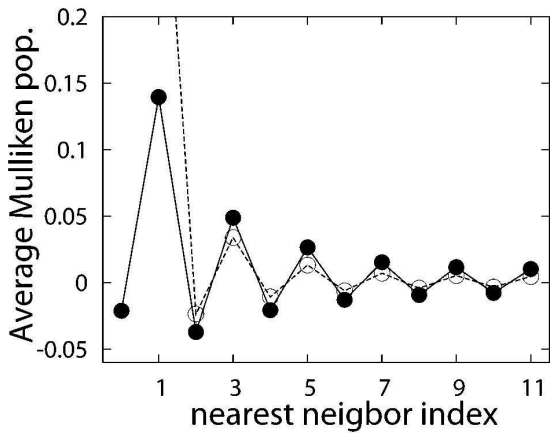}
\caption{\label{fig:eps5} Dependence of the spin populations 
averaged over \textit{i}th nearest neighbors of the hydrogen 
chemisorption defects ($\bullet$) and the vacancy defects ($\circ$). 
The spin population for the first nearest neighbor atoms of 
the vacancy defect (0.39) are out of scale due to the contribution 
of the localized $sp^2$ dangling bond state.}
\end{figure}

The distributions of the electron spin magnetization density in 
the vicinity of both types of defects clearly show the characteristic 
$\sqrt{3}$$\times$$\sqrt{3}$ patterns also observed for the charge density
in the STM experiments.
For the hydrogen chemisorption defect the projection of 
the spin density (Fig.~\ref{fig:eps4}a) on the graphene plane 
clearly shows three-fold symmetry. For the vacancy type
defect the symmetry is broken due to the Jahn-Teller 
distortion (Fig.~\ref{fig:eps4}b). 
The localized magnetic moment associated to the dangling bond 
of atom 1 can also be observed. 
The simulated STM images (Fig.~\ref{fig:eps4}c and \ref{fig:eps4}d) based on 
our calculations agree with experimental observations.\cite{Mizes89,Ruffieux00,Ruffieux05}
The distribution of the electron spin density is represented 
in Fig.~\ref{fig:eps5} ($n$=6 model) by means of the Mulliken 
spin populations averaged over \textit{i}th nearest neighbors 
to the defect atom. 
The spin populations show a damped oscillation behavior 
as a function of the nearest neighbor index and, therefore, 
of the distance to the defect.  
The magnetization pattern is explained by the fact that the 
defect state is distributed over the sites of the sublattice 
complementary to the one in which the 
defect was created (i. e. over the odd nearest neighbors), 
and shows a power law decay.\cite{Pereira06}
The major positive contribution to the electron spin density 
is defined by the exchange splitting of the defect states. 
In addition, the exchange spin-polarization effect (i.e. 
the response of the fully populated valence bands to 
the magnetization of the defect states) results in a negative 
spin density on the even nearest neighbor sites 
and in the enhancement of a positive spin density on the odd 
nearest neighbor sites (see Fig.~\ref{fig:eps4}). 
A similar phenomenon takes place in the case of the neutral bond length 
alternation defect states 
(neutral solitons) 
in one-dimensional 
polyene chains.\cite{Thomann83,Kirtman91}
The calculated magnitude of the negative spin-polarization is 
$\approx$$1/3$ of the positive spin populations on the neighbor 
sites in the vicinity of the defect site. 
This is close to the ratio observed for the {\it trans}-polyacetylene.\cite{Thomann83}
The magnitudes of the spin populations are lower in the case 
of the vacancy defect because of the fractional spin-polarization 
of the defect band. 

\begin{figure}
\includegraphics[width=7.5cm]{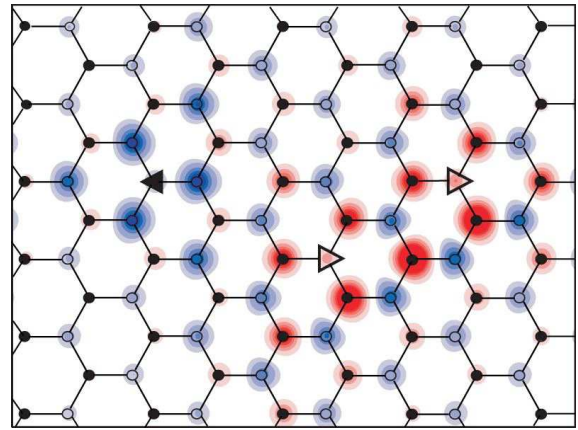}
\caption{\label{fig:eps6} (Color online) 
Spin density distribution in the system with three
hydrogen chemisorption defects (two defects in sublattice $\alpha$
($\triangleright$) and one in sublattice $\beta$ ($\blacktriangleleft$)).
The same scale as in Fig.~\ref{fig:eps4} applies.}
\end{figure}

According to the Stoner picture, the magnetic ordering is driven 
by the exchange energy $E_x \sim -\sum_i M^2_j$
with $M_j$ being the magnetization of the $p_z$ orbital of the 
$j$th carbon atom.\cite{Mohn03} 
Ferromagnetic ordering is the only possibility for the magnetism 
originating from quasilocalized states induced by defects in the
same sublattice because of the non-oscillating behavior of both 
(i) $M_j$ within the same sublattice  
and (ii) the indirect (RKKY) coupling due to the 
semimetallic properties of graphene.\cite{Vozmediano05}
On the contrary, for the case of defect states in different sublattices, 
$E_x$ is minimized when the coupling is antiferromagnetic. In this case, 
the mechanism of the exchange coupling is defined by the indirect 
spin-polarization effect. 
The strength of the coupling between the defect-induced magnetic moments 
located in different sublattices depends on the defect 
concentration since $E_x \sim -\sum_i M^2_j$. 
The contribution of the magnetic moment associated with a single defect is
$\sum_i M^2_j \sim \sum_j |\psi_d(\vec{r}_j)|^4 \sim log^{-2}(N)$, where $1/N$
is the defect concentration.\cite{Pereira06}
To further illustrate this point, 
we calculated the ground state magnetic configuration of the 
system with three close hydrogen chemisorption defects using the 
DFT approach. 
We found that in the ground state configuration two defects in 
the $\alpha$ sublattice are coupled ferromagnetically with each other
and antiferromagnetically with the third defect 
in the $\beta$ sublattice (Fig.~\ref{fig:eps6}).
The resulting magnetic moment of this system amounts to 2$-$1=1~$\mu_B$
in agreement with the second Lieb theorem,\cite{Lieb89} and the characteristic $\sqrt{3}$$\times$$\sqrt{3}$ patterns 
of the magnetization density associated with individual defects 
can be recognized. In nanographite materials with defects present 
with an equal probability in both sublattices, 
the overall correlation of the magnetic moments is expected 
to be antiferromagnetic. The antiferromagnetic ordering was 
experimentally observed in carbon nanohorns.\cite{Garaj00,Imai06}

In addition, we found that the proton spin of chemisorption hydrogen atom 
shows a strong magnetic coupling to the electron in 
the quasilocalized defect state. 
Even for the lowest defect concentration 
studied we find the magnitude of Fermi contact hyperfine interaction of $\approx$35~Gauss. 
One can consider the nuclear spin quantum computer
implementation originally proposed by Kane\cite{Kane98} 
based on carbon nanostructures.
   
\section{CONCLUSIONS}
   
In conclusion, our calculations reveal the itinerant 
magnetism triggered by simple defects in graphene and stable over the
wide range of concentrations. 
It is notable, that the itinerant magnetism does not require the presence of 
highly reactive unsaturated dangling bonds.
Both ferromagnetic and antiferromagnetic scenarios of the magnetic correlation 
are possible with the second being more probable for truly disordered systems.
The reconstruction of vacancy defects was found to be responsible for the
partial suppression of magnetic moments and for the broadening of defect bands. 
We argue that the defect-induced itinerant magnetism is responsible for 
the experimentally observed high-$T_C$ ferromagnetism of irradiated graphite samples.

\section*{ACKNOWLEDGMENTS}

Authors acknowledge G. Buchs, D. Ivanov, M. I. Katsnelson 
and I. Tavernelli for discussions.
O.~Y. thanks the Swiss National Science Foundation for financial support. 
The computational resources were provided by 
the Swiss Center for Scientific Computing and the DIT-EPFL.


\begin{thebibliography}{31}

\expandafter\ifx\csname natexlab\endcsname\relax\def\natexlab#1{#1}\fi
\expandafter\ifx\csname bibnamefont\endcsname\relax
  \def\bibnamefont#1{#1}\fi
\expandafter\ifx\csname bibfnamefont\endcsname\relax
  \def\bibfnamefont#1{#1}\fi
\expandafter\ifx\csname citenamefont\endcsname\relax
  \def\citenamefont#1{#1}\fi
\expandafter\ifx\csname url\endcsname\relax
  \def\url#1{\texttt{#1}}\fi
\expandafter\ifx\csname urlprefix\endcsname\relax\def\urlprefix{URL }\fi
\providecommand{\bibinfo}[2]{#2}
\providecommand{\eprint}[2][]{\url{#2}}

\bibitem[{\citenamefont{Makarova and Palacio}(2005)}]{Makarova05}
\bibinfo{editor}{\bibfnamefont{T.}~\bibnamefont{Makarova}} \bibnamefont{and}
  \bibinfo{editor}{\bibfnamefont{F.}~\bibnamefont{Palacio}}, eds.,
  \emph{\bibinfo{title}{Carbon-Based Magnetism: An Overview of Metal Free
  Carbon-Based Compounds and Materials}} (\bibinfo{publisher}{Elsevier},
  \bibinfo{address}{Amsterdam}, \bibinfo{year}{2005}).

\bibitem[{\citenamefont{Coey et~al.}(2002)\citenamefont{Coey, Venkatesan,
  Fitzgerald, Douvalis, and Sanders}}]{Coey02}
\bibinfo{author}{\bibfnamefont{J.~M.~D.} \bibnamefont{Coey}},
  \bibinfo{author}{\bibfnamefont{M.}~\bibnamefont{Venkatesan}},
  \bibinfo{author}{\bibfnamefont{C.~B.} \bibnamefont{Fitzgerald}},
  \bibinfo{author}{\bibfnamefont{A.~P.} \bibnamefont{Douvalis}},
  \bibnamefont{and} \bibinfo{author}{\bibfnamefont{I.~S.}
  \bibnamefont{Sanders}}, 
  \bibinfo{journal}{Nature}
  \textbf{\bibinfo{volume}{420}}, \bibinfo{pages}{156} (\bibinfo{year}{2002}).

\bibitem[{\citenamefont{Lee et~al.}(2005)\citenamefont{Lee, Son, Park, Han, and
  Yu}}]{Lee05}
\bibinfo{author}{\bibfnamefont{H.}~\bibnamefont{Lee}},
  \bibinfo{author}{\bibfnamefont{Y.-W.} \bibnamefont{Son}},
  \bibinfo{author}{\bibfnamefont{N.}~\bibnamefont{Park}},
  \bibinfo{author}{\bibfnamefont{S.}~\bibnamefont{Han}}, \bibnamefont{and}
  \bibinfo{author}{\bibfnamefont{J.}~\bibnamefont{Yu}}, 
  \bibinfo{journal}{Phys.
  Rev. B} \textbf{\bibinfo{volume}{72}}, \bibinfo{pages}{174431}
  (\bibinfo{year}{2005}).

\bibitem[{\citenamefont{Okada and Oshiyama}(2001)}]{Okada01}
\bibinfo{author}{\bibfnamefont{S.}~\bibnamefont{Okada}} \bibnamefont{and}
  \bibinfo{author}{\bibfnamefont{A.}~\bibnamefont{Oshiyama}},
  \bibinfo{journal}{Phys. Rev. Lett.} \textbf{\bibinfo{volume}{87}},
  \bibinfo{pages}{146803} (\bibinfo{year}{2001}).

\bibitem[{\citenamefont{Kim et~al.}(2003)\citenamefont{Kim, amd K.~J.~Chang,
  and Tom\'anek}}]{Kim03}
\bibinfo{author}{\bibfnamefont{Y.-H.} \bibnamefont{Kim}}, 
  \bibinfo{author}{\bibfnamefont{J.~Choi}},
  \bibinfo{author}{\bibnamefont{K.~J.~Chang}},
  \bibnamefont{and}
  \bibinfo{author}{\bibfnamefont{D.}~\bibnamefont{Tom\'anek}},
  \bibinfo{journal}{Phys. Rev. B} \textbf{\bibinfo{volume}{68}},
  \bibinfo{pages}{125420} (\bibinfo{year}{2003}).

\bibitem[{\citenamefont{Lehtinen et~al.}(2004)\citenamefont{Lehtinen, Foster,
  Ma, Krasheninnikov, and Nieminen}}]{Lehtinen04}
\bibinfo{author}{\bibfnamefont{P.~O.}~\bibnamefont{Lehtinen}}, 
  \bibinfo{author}{\bibfnamefont{A.~S.} \bibnamefont{Foster}},
  \bibinfo{author}{\bibfnamefont{Y.}~\bibnamefont{Ma}},
  \bibinfo{author}{\bibfnamefont{A.~V.}~\bibnamefont{Krasheninnikov}},
  \bibnamefont{and} \bibinfo{author}{\bibfnamefont{R.~M.}
  \bibnamefont{Nieminen}}, 
  \bibinfo{journal}{Phys. Rev. Lett.}
  \textbf{\bibinfo{volume}{93}}, \bibinfo{pages}{187202}
  (\bibinfo{year}{2004}).

\bibitem[{\citenamefont{Duplock et~al.}(2004)\citenamefont{Duplock, Scheffler,
  and Lindan}}]{Duplock04}
\bibinfo{author}{\bibfnamefont{E.~J.} \bibnamefont{Duplock}},
  \bibinfo{author}{\bibfnamefont{M.}~\bibnamefont{Scheffler}},
  \bibnamefont{and} \bibinfo{author}{\bibfnamefont{P.~J.~D.}
  \bibnamefont{Lindan}}, \bibinfo{journal}{Phys. Rev. Lett.}
  \textbf{\bibinfo{volume}{92}}, \bibinfo{pages}{225502}
  (\bibinfo{year}{2004}).
  
\bibitem[{\citenamefont{Carlsson and Scheffler}(2006)}]{Carlsson06}
\bibinfo{author}{\bibfnamefont{J.~M.} \bibnamefont{Carlsson}} \bibnamefont{and}
  \bibinfo{author}{\bibfnamefont{M.} \bibnamefont{Scheffler}},
  \bibinfo{journal}{Phys. Rev. Lett.} \textbf{\bibinfo{volume}{96}},
  \bibinfo{pages}{046806} (\bibinfo{year}{2006}).

\bibitem[{\citenamefont{Charlier}(2002)}]{Charlier02}
\bibinfo{author}{\bibfnamefont{J.-C.} \bibnamefont{Charlier}},
  \bibinfo{journal}{Acc. Chem. Res.} \textbf{\bibinfo{volume}{35}},
  \bibinfo{pages}{1063} (\bibinfo{year}{2002}).

\bibitem[{\citenamefont{Ruffieux et~al.}(2000)\citenamefont{Ruffieux,
  Gr\"oning, Schwaller, Schlapbach, and Gr\"oning}}]{Ruffieux00}
\bibinfo{author}{\bibfnamefont{P.}~\bibnamefont{Ruffieux}}, 
  \bibinfo{author}{\bibfnamefont{O.}~\bibnamefont{Gr\"oning}},
  \bibinfo{author}{\bibfnamefont{P.}~\bibnamefont{Schwaller}},
  \bibinfo{author}{\bibfnamefont{L.}~\bibnamefont{Schlapbach}},
  \bibnamefont{and}
  \bibinfo{author}{\bibfnamefont{P.}~\bibnamefont{Gr\"oning}},
  \bibinfo{journal}{Phys. Rev. Lett.} \textbf{\bibinfo{volume}{84}},
  \bibinfo{pages}{4910} (\bibinfo{year}{2000}).

\bibitem[{\citenamefont{Esquinazi et~al.}(2003)\citenamefont{Esquinazi,
  Spemann, H\"ohne, Setzer, Han, and Butz}}]{Esquinazi03}
\bibinfo{author}{\bibfnamefont{P.}~\bibnamefont{Esquinazi}}, 
  \bibinfo{author}{\bibfnamefont{D.}~\bibnamefont{Spemann}},
  \bibinfo{author}{\bibfnamefont{R.}~\bibnamefont{H\"ohne}},
  \bibinfo{author}{\bibfnamefont{A.}~\bibnamefont{Setzer}},
  \bibinfo{author}{\bibfnamefont{K.-H.} \bibnamefont{Han}}, \bibnamefont{and}
  \bibinfo{author}{\bibfnamefont{T.}~\bibnamefont{Butz}},
  \bibinfo{journal}{Phys. Rev. Lett.} \textbf{\bibinfo{volume}{91}},
  \bibinfo{pages}{227201} (\bibinfo{year}{2003}).

\bibitem[{\citenamefont{Hashimoto et~al.}(2004)\citenamefont{Hashimoto,
  Suenaga, Gloter, Urita, and Iijima}}]{Hashimoto04}
\bibinfo{author}{\bibfnamefont{A.}~\bibnamefont{Hashimoto}}, 
  \bibinfo{author}{\bibfnamefont{K.}~\bibnamefont{Suenaga}},
  \bibinfo{author}{\bibfnamefont{A.}~\bibnamefont{Gloter}},
  \bibinfo{author}{\bibfnamefont{K.}~\bibnamefont{Urita}}, \bibnamefont{and}
  \bibinfo{author}{\bibfnamefont{S.}~\bibnamefont{Iijima}},
  \bibinfo{journal}{Nature} \textbf{\bibinfo{volume}{430}},
  \bibinfo{pages}{870} (\bibinfo{year}{2004}).

\bibitem[{\citenamefont{Urita et~al.}(2005)\citenamefont{Urita, Suenaga, Sugai,
  Shinohara, and Iijima}}]{Urita05}
\bibinfo{author}{\bibfnamefont{K.}~\bibnamefont{Urita}}, 
  \bibinfo{author}{\bibfnamefont{K.}~\bibnamefont{Suenaga}},
  \bibinfo{author}{\bibfnamefont{T.}~\bibnamefont{Sugai}},
  \bibinfo{author}{\bibfnamefont{H.}~\bibnamefont{Shinohara}},
  \bibnamefont{and} \bibinfo{author}{\bibfnamefont{S.}~\bibnamefont{Iijima}},
  \bibinfo{journal}{Phys. Rev. Lett.} \textbf{\bibinfo{volume}{94}},
  \bibinfo{pages}{155502} (\bibinfo{year}{2005}).

\bibitem[{\citenamefont{Banhart}(1999)}]{Banhart99}
\bibinfo{author}{\bibfnamefont{F.}~\bibnamefont{Banhart}},
  \bibinfo{journal}{Rep. Prog. Phys.} \textbf{\bibinfo{volume}{62}},
  \bibinfo{pages}{1181} (\bibinfo{year}{1999}).

\bibitem[{\citenamefont{Mik\'o et~al.}(2003)\citenamefont{Mik\'o, Milas, Seo,
  Couteau, Barisic, Ga\'al, and Forr\'o}}]{Miko03}
\bibinfo{author}{\bibfnamefont{C.}~\bibnamefont{Mik\'o}}, 
  \bibinfo{author}{\bibfnamefont{M.}~\bibnamefont{Milas}},
  \bibinfo{author}{\bibfnamefont{J.~W.} \bibnamefont{Seo}},
  \bibinfo{author}{\bibfnamefont{E.}~\bibnamefont{Couteau}},
  \bibinfo{author}{\bibfnamefont{N.}~\bibnamefont{Barisic}},
  \bibinfo{author}{\bibfnamefont{R.}~\bibnamefont{Ga\'al}}, \bibnamefont{and}
  \bibinfo{author}{\bibfnamefont{L.}~\bibnamefont{Forr\'o}},
  \bibinfo{journal}{Appl. Phys. Lett.} \textbf{\bibinfo{volume}{83}},
  \bibinfo{pages}{4622} (\bibinfo{year}{2003}).

\bibitem[{\citenamefont{Han et~al.}(2003)\citenamefont{Han, Stepmann,
  Esquinazi, H\:ohne, Riede, and Butz}}]{Han03}
\bibinfo{author}{\bibfnamefont{K.-H.} \bibnamefont{Han}}, 
  \bibinfo{author}{\bibfnamefont{D.}~\bibnamefont{Stepmann}},
  \bibinfo{author}{\bibfnamefont{P.}~\bibnamefont{Esquinazi}},
  \bibinfo{author}{\bibfnamefont{R.}~\bibnamefont{H\"ohne}},
  \bibinfo{author}{\bibfnamefont{V.}~\bibnamefont{Riede}}, \bibnamefont{and}
  \bibinfo{author}{\bibfnamefont{T.}~\bibnamefont{Butz}},
  \bibinfo{journal}{Adv. Mater.} \textbf{\bibinfo{volume}{15}},
  \bibinfo{pages}{1719} (\bibinfo{year}{2003}).

\bibitem[{\citenamefont{Kis et~al.}(2004)\citenamefont{Kis, Cs\'anyi, Salvetat,
  Lee, Couteau, Kulik, Benoit, Brugger, and Forr\'o}}]{Kis04}
\bibinfo{author}{\bibfnamefont{A.}~\bibnamefont{Kis}}, 
  \bibinfo{author}{\bibfnamefont{G.}~\bibnamefont{Cs\'anyi}},
  \bibinfo{author}{\bibfnamefont{J.-P.} \bibnamefont{Salvetat}},
  \bibinfo{author}{\bibfnamefont{T.-N.} \bibnamefont{Lee}},
  \bibinfo{author}{\bibfnamefont{E.}~\bibnamefont{Couteau}},
  \bibinfo{author}{\bibfnamefont{A.~J.} \bibnamefont{Kulik}},
  \bibinfo{author}{\bibfnamefont{W.}~\bibnamefont{Benoit}},
  \bibinfo{author}{\bibfnamefont{J.}~\bibnamefont{Brugger}}, \bibnamefont{and}
  \bibinfo{author}{\bibfnamefont{L.}~\bibnamefont{Forr\'o}},
  \bibinfo{journal}{Nat. Mater.} \textbf{\bibinfo{volume}{3}},
  \bibinfo{pages}{153} (\bibinfo{year}{2004}).

\bibitem[{\citenamefont{Mizes and Foster}(1989)}]{Mizes89}
\bibinfo{author}{\bibfnamefont{H.~A.} \bibnamefont{Mizes}} \bibnamefont{and}
  \bibinfo{author}{\bibfnamefont{J.~S.} \bibnamefont{Foster}},
  \bibinfo{journal}{Science} \textbf{\bibinfo{volume}{244}},
  \bibinfo{pages}{559} (\bibinfo{year}{1989}).

\bibitem[{\citenamefont{Ruffieux et~al.}(2005)\citenamefont{Ruffieux,
  Melle-Franco, Gr\"oning, Bielmann, Zerbetto, and Gr\"oning}}]{Ruffieux05}
\bibinfo{author}{\bibfnamefont{P.}~\bibnamefont{Ruffieux}}, 
  \bibinfo{author}{\bibfnamefont{M.}~\bibnamefont{Melle-Franco}},
  \bibinfo{author}{\bibfnamefont{O.}~\bibnamefont{Gr\"oning}},
  \bibinfo{author}{\bibfnamefont{M.}~\bibnamefont{Bielmann}},
  \bibinfo{author}{\bibfnamefont{F.}~\bibnamefont{Zerbetto}}, \bibnamefont{and}
  \bibinfo{author}{\bibfnamefont{P.}~\bibnamefont{Gr\"oning}},
  \bibinfo{journal}{Phys. Rev. B} \textbf{\bibinfo{volume}{71}},
  \bibinfo{pages}{153403} (\bibinfo{year}{2005}).

\bibitem[{\citenamefont{Pereira et~al.}(2006)\citenamefont{Pereira, Guinea,
  Lopes~dos Santos, Peres, and Neto}}]{Pereira06}
\bibinfo{author}{\bibfnamefont{V.~M.} \bibnamefont{Pereira}},
  \bibinfo{author}{\bibfnamefont{F.}~\bibnamefont{Guinea}},
  \bibinfo{author}{\bibfnamefont{J.~M.~B.} \bibnamefont{Lopes~dos Santos}},
  \bibinfo{author}{\bibfnamefont{N.~M.~R.} \bibnamefont{Peres}},
  \bibnamefont{and} \bibinfo{author}{\bibfnamefont{A.~H.}
  \bibnamefont{Castro Neto}}, \bibinfo{journal}{Phys. Rev. Lett.}
  \textbf{\bibinfo{volume}{96}}, \bibinfo{pages}{036801}
  (\bibinfo{year}{2006}).

\bibitem[{\citenamefont{Kelly and Halas}(1998)}]{Kelly98}
\bibinfo{author}{\bibfnamefont{K.}~\bibnamefont{Kelly}} \bibnamefont{and}
  \bibinfo{author}{\bibfnamefont{N.}~\bibnamefont{Halas}},
  \bibinfo{journal}{Surf. Sci.} \textbf{\bibinfo{volume}{416}},
  \bibinfo{pages}{L1085} (\bibinfo{year}{1998}).
  
\bibitem[{\citenamefont{Edwards and Katsnelson}(2006)}]{Edwards06}
\bibinfo{author}{\bibfnamefont{D.~M.} \bibnamefont{Edwards}} and 
  \bibinfo{author}{\bibfnamefont{M.~I.} \bibnamefont{Katsnelson}},
  \bibinfo{journal}{J. Phys.: Condens. Matter} \textbf{\bibinfo{volume}{18}},
  \bibinfo{pages}{7209} (\bibinfo{year}{2006}).
  
\bibitem[{\citenamefont{RKKY}(2006)}]{RKKY}
\bibinfo{author}{\bibfnamefont{M.~A.} \bibnamefont{Rudermann}} and 
  \bibinfo{author}{\bibfnamefont{C.} \bibnamefont{Kittel}},
  \bibinfo{journal}{Phys. Rev.} \textbf{\bibinfo{volume}{96}},
  \bibinfo{pages}{99} (\bibinfo{year}{1954});
\bibinfo{author}{\bibfnamefont{T.} \bibnamefont{Kasuya}},
  \bibinfo{journal}{Progr. Theor. Phys. (Japan)} \textbf{\bibinfo{volume}{16}},
  \bibinfo{pages}{45} (\bibinfo{year}{1965});
\bibinfo{author}{\bibfnamefont{K.} \bibnamefont{Yoshida}},
  \bibinfo{journal}{Phys. Rev.} \textbf{\bibinfo{volume}{106}},
  \bibinfo{pages}{893} (\bibinfo{year}{1958}).
    
\bibitem[{\citenamefont{Vozmediano et~al.}(2005)\citenamefont{Vozmediano,
  L\'opez-Sancho, Stauber, and Guinea}}]{Vozmediano05}
\bibinfo{author}{\bibfnamefont{M.~A.~H.} \bibnamefont{Vozmediano}}, 
  \bibinfo{author}{\bibfnamefont{M.~P.} \bibnamefont{L\'opez-Sancho}},
  \bibinfo{author}{\bibfnamefont{T.}~\bibnamefont{Stauber}}, \bibnamefont{and}
  \bibinfo{author}{\bibfnamefont{F.}~\bibnamefont{Guinea}},
  \bibinfo{journal}{Phys.\ Rev.\ B} \textbf{\bibinfo{volume}{72}},
  \bibinfo{pages}{155121} (\bibinfo{year}{2005}).

\bibitem[{\citenamefont{Dugaev}(2006)}]{Dugaev06}
\bibinfo{author}{\bibfnamefont{V.~K.} \bibnamefont{Dugaev}},
  \bibinfo{author}{\bibfnamefont{V.~I.} \bibnamefont{Litvinov}},  and 
  \bibinfo{author}{\bibfnamefont{J.} \bibnamefont{Barnas}},
  \bibinfo{journal}{Phys. Rev. B} \textbf{\bibinfo{volume}{74}},
  \bibinfo{pages}{224438} (\bibinfo{year}{2006}).

\bibitem[{\citenamefont{Artacho et~al.}(2004)\citenamefont{Artacho, Gale,
  Garc\'ia, Junquera, Martin, Ordej\'on, S\'anchez-Portal, and Soler}}]{SIESTA}
\bibinfo{author}{\bibfnamefont{E.}~\bibnamefont{Artacho}} {\it et al.},
  \bibinfo{author}{\bibfnamefont{J.~D.} \bibnamefont{Gale}},
  \bibinfo{author}{\bibfnamefont{A.}~\bibnamefont{Garc\'ia}},
  \bibinfo{author}{\bibfnamefont{J.}~\bibnamefont{Junquera}},
  \bibinfo{author}{\bibfnamefont{R.~M.} \bibnamefont{Martin}},
  \bibinfo{author}{\bibfnamefont{P.}~\bibnamefont{Ordej\'on}},
  \bibinfo{author}{\bibfnamefont{D.}~\bibnamefont{S\'anchez-Portal}},
  \bibnamefont{and} \bibinfo{author}{\bibfnamefont{J.~M.} \bibnamefont{Soler}},
   \texttt{SIESTA}, \textit{version 1.3} (\bibinfo{year}{2004}).

\bibitem[{\citenamefont{Soler et~al.}(2002)\citenamefont{Soler, Artacho, Gale,
  Garc\'ia, Junquera, Ordej\'on, and S\'anchez-Portal}}]{Soler02}
\bibinfo{author}{\bibfnamefont{J.~M.} \bibnamefont{Soler}}, 
  \bibinfo{author}{\bibfnamefont{E.}~\bibnamefont{Artacho}},
  \bibinfo{author}{\bibfnamefont{J.~D.} \bibnamefont{Gale}},
  \bibinfo{author}{\bibfnamefont{A.}~\bibnamefont{Garc\'ia}},
  \bibinfo{author}{\bibfnamefont{J.}~\bibnamefont{Junquera}},
  \bibinfo{author}{\bibfnamefont{P.}~\bibnamefont{Ordej\'on}},
  \bibnamefont{and}
  \bibinfo{author}{\bibfnamefont{D.}~\bibnamefont{S\'anchez-Portal}},
  \bibinfo{journal}{J. Phys.: Condens. Matter} \textbf{\bibinfo{volume}{14}},
  \bibinfo{pages}{2745} (\bibinfo{year}{2002}).

\bibitem[{\citenamefont{Perdew et~al.}(1996)\citenamefont{Perdew, Burke, and
  Ernzerhof}}]{Perdew96}
\bibinfo{author}{\bibfnamefont{J.~P.} \bibnamefont{Perdew}},
  \bibinfo{author}{\bibfnamefont{K.}~\bibnamefont{Burke}}, \bibnamefont{and}
  \bibinfo{author}{\bibfnamefont{M.}~\bibnamefont{Ernzerhof}},
  \bibinfo{journal}{Phys. Rev. Lett.} \textbf{\bibinfo{volume}{77}},
  \bibinfo{pages}{3865} (\bibinfo{year}{1996}).

\bibitem[{\citenamefont{Monkhorst and Pack}(1976)}]{Monkhorst76}
\bibinfo{author}{\bibfnamefont{H.~J.} \bibnamefont{Monkhorst}}
  \bibnamefont{and} \bibinfo{author}{\bibfnamefont{J.~D.} \bibnamefont{Pack}},
  \bibinfo{journal}{Phys. Rev. B} \textbf{\bibinfo{volume}{16}},
  \bibinfo{pages}{5188} (\bibinfo{year}{1976}).

\bibitem[{\citenamefont{Ruffieux et~al.}(2002)\citenamefont{Ruffieux,
  Gr\"oning, Bielmann, Mauron, Schlapbach, and Gr\"oning}}]{Ruffieux02}
\bibinfo{author}{\bibfnamefont{P.}~\bibnamefont{Ruffieux}}, 
  \bibinfo{author}{\bibfnamefont{O.}~\bibnamefont{Gr\"oning}},
  \bibinfo{author}{\bibfnamefont{M.}~\bibnamefont{Bielmann}},
  \bibinfo{author}{\bibfnamefont{P.}~\bibnamefont{Mauron}},
  \bibinfo{author}{\bibfnamefont{L.}~\bibnamefont{Schlapbach}},
  \bibnamefont{and}
  \bibinfo{author}{\bibfnamefont{P.}~\bibnamefont{Gr\"oning}},
  \bibinfo{journal}{Phys. Rev. B} \textbf{\bibinfo{volume}{66}},
  \bibinfo{pages}{245416} (\bibinfo{year}{2002}).

\bibitem[{\citenamefont{Lu and Pan}(2004)}]{Lu04}
\bibinfo{author}{\bibfnamefont{A.~J.} \bibnamefont{Lu}} \bibnamefont{and}
  \bibinfo{author}{\bibfnamefont{B.~C.} \bibnamefont{Pan}},
  \bibinfo{journal}{Phys. Rev. Lett.} \textbf{\bibinfo{volume}{92}},
  \bibinfo{pages}{105504} (\bibinfo{year}{2004}).

\bibitem[{\citenamefont{Mohn}(2003)}]{Mohn03}
\bibinfo{author}{\bibfnamefont{P.}~\bibnamefont{Mohn}},
  \emph{\bibinfo{title}{Magnetism in the solid state}}
  (\bibinfo{publisher}{Springer-Verlag}, \bibinfo{address}{Berlin Heidelberg},
  \bibinfo{year}{2003}).

\bibitem[{\citenamefont{Thomann et~al.}(1983)\citenamefont{Thomann, Dalton,
  Tomkiewicz, Shiren, and Clarke}}]{Thomann83}
\bibinfo{author}{\bibfnamefont{H.}~\bibnamefont{Thomann}}, 
  \bibinfo{author}{\bibfnamefont{L.~R.} \bibnamefont{Dalton}},
  \bibinfo{author}{\bibfnamefont{Y.}~\bibnamefont{Tomkiewicz}},
  \bibinfo{author}{\bibfnamefont{N.~S.} \bibnamefont{Shiren}},
  \bibnamefont{and} \bibinfo{author}{\bibfnamefont{T.~C.}
  \bibnamefont{Clarke}}, 
\bibinfo{journal}{Phys. Rev. Lett.}
  \textbf{\bibinfo{volume}{50}}, \bibinfo{pages}{533} (\bibinfo{year}{1983}).

\bibitem[{\citenamefont{Kirtman et~al.}(1991)\citenamefont{Kirtman, Hasan, and
  Chipman}}]{Kirtman91}
\bibinfo{author}{\bibfnamefont{B.}~\bibnamefont{Kirtman}},
  \bibinfo{author}{\bibfnamefont{M.}~\bibnamefont{Hasan}}, \bibnamefont{and}
  \bibinfo{author}{\bibfnamefont{D.~M.} \bibnamefont{Chipman}},
  \bibinfo{journal}{J. Chem. Phys.} \textbf{\bibinfo{volume}{95}},
  \bibinfo{pages}{7698} (\bibinfo{year}{1991}).
  
\bibitem[{\citenamefont{Lieb}(1989)}]{Lieb89}
\bibinfo{author}{\bibfnamefont{E.~H.} \bibnamefont{Lieb}},
  \bibinfo{journal}{Phys. Rev. Lett.} \textbf{\bibinfo{volume}{62}},
  \bibinfo{pages}{1201} (\bibinfo{year}{1989})
[Erratum: \bibinfo{journal}{Phys. Rev. Lett.}, \textbf{\bibinfo{volume}{62}}, \bibinfo{pages}{1927} (\bibinfo{year}{1989})].

\bibitem[{\citenamefont{Garaj et~al.}(2000)\citenamefont{Garaj, Thien-Nga,
  Gaal, Forro´, Takahashi, Kokai, Yudasaka, and Iijima}}]{Garaj00}
\bibinfo{author}{\bibfnamefont{S.}~\bibnamefont{Garaj}}, 
  \bibinfo{author}{\bibfnamefont{L.}~\bibnamefont{Thien-Nga}},
  \bibinfo{author}{\bibfnamefont{R.}~\bibnamefont{Gaal}},
  \bibinfo{author}{\bibfnamefont{L.}~\bibnamefont{Forro´}},
  \bibinfo{author}{\bibfnamefont{K.}~\bibnamefont{Takahashi}},
  \bibinfo{author}{\bibfnamefont{F.}~\bibnamefont{Kokai}},
  \bibinfo{author}{\bibfnamefont{M.}~\bibnamefont{Yudasaka}}, \bibnamefont{and}
  \bibinfo{author}{\bibfnamefont{S.}~\bibnamefont{Iijima}},
  \bibinfo{journal}{Phys. Rev. B} \textbf{\bibinfo{volume}{62}},
  \bibinfo{pages}{17115} (\bibinfo{year}{2000}).

\bibitem[{\citenamefont{Imai et~al.}(2006)\citenamefont{Imai, Babu, Oldfield,
  Wieckowski, Kasuya, Azami, Shimakawa, Yudasaka, Kubo, and Iijima}}]{Imai06}
\bibinfo{author}{\bibfnamefont{H.}~\bibnamefont{Imai}}, 
  \bibinfo{author}{\bibfnamefont{P.~K.} \bibnamefont{Babu}},
  \bibinfo{author}{\bibfnamefont{E.}~\bibnamefont{Oldfield}},
  \bibinfo{author}{\bibfnamefont{A.}~\bibnamefont{Wieckowski}},
  \bibinfo{author}{\bibfnamefont{D.}~\bibnamefont{Kasuya}},
  \bibinfo{author}{\bibfnamefont{T.}~\bibnamefont{Azami}},
  \bibinfo{author}{\bibfnamefont{Y.}~\bibnamefont{Shimakawa}},
  \bibinfo{author}{\bibfnamefont{M.}~\bibnamefont{Yudasaka}},
  \bibinfo{author}{\bibfnamefont{Y.}~\bibnamefont{Kubo}}, \bibnamefont{and}
  \bibinfo{author}{\bibfnamefont{S.}~\bibnamefont{Iijima}},
  \bibinfo{journal}{Phys. Rev. B} \textbf{\bibinfo{volume}{73}},
  \bibinfo{pages}{125405} (\bibinfo{year}{2006}).

\bibitem[{\citenamefont{Kane}(1998)}]{Kane98}
\bibinfo{author}{\bibfnamefont{B.~E.} \bibnamefont{Kane}},
  \bibinfo{journal}{Nature} \textbf{\bibinfo{volume}{393}},
  \bibinfo{pages}{133} (\bibinfo{year}{1998}).


\end{thebibliography}

\end{document}